\documentclass[preprint,12pt]{elsarticle}

\usepackage{graphicx}
\usepackage{makeidx}
\usepackage[]{hyperref}

\usepackage{amsmath}
\usepackage{amssymb,epsfig,graphics}
\usepackage{amscd}
\usepackage{verbatim}

\usepackage{amsmath}
\usepackage{amsthm}
\usepackage{amsfonts}

\begin{document}

\begin{frontmatter}

\title{Calculation of the electronic structure near the tip of a graphitic nanocone}

\author[aff1, aff2]{R. Pincak\corref{cor1}}\ead{pincak@saske.sk}
\author[aff2, aff3]{J. Smotlacha\corref{cor1}}\ead{smota@centrum.cz}
\author[aff1]{M.Pudlak}\ead{pudlak@saske.sk}

\cortext[cor1]{Phone numbers: +7 925 1146637 (J. Smotlacha), +421 0907 936147 (R. Pincak)}

\address[aff1]{Institute of Experimental Physics, Slovak Academy of Sciences,
Watsonova 47,043 53 Kosice, Slovak Republic}
\address[aff2]{Bogoliubov Laboratory of Theoretical Physics, Joint
Institute for Nuclear Research, 141980 Dubna, Moscow region, Russia}
\address[aff3]{Faculty of Nuclear Sciences and Physical Engineering, Czech Technical University, Brehova 7, 110 00 Prague,
Czech Republic}

\date{\today}


\begin{abstract}
In the earlier works, the electronic structure of the graphitic
nanocone for the long distance from the tip was investigated. Here,
we investigate the behaviour of the given nanostructure near the tip
where in our approach hybridizations of $\pi$ orbitals need to be
included. In this case, the curvature dependence of $\pi$ orbital
energy has to be imposed into the model. For this purpose, we use an
approximation valid for small values of the corresponding
parameters. We consider different numbers of the pentagonal defects
in the tip. This localization of the electrons on the nanocone tip could be used as a real application in the electron microscopes.
\end{abstract}

\end{frontmatter}

\section{Introduction}

In recent years, a lot of nanostructural surfaces was investigated.
The interest in them is due to their electronic and transport properties.
They can be used as nanoscale devices like transistors, molecular memory devices, nanowires, etc.
They can be produced in complex thermal processes such as, for example, the chemical vapor deposition.

The electronic properties are mediated by different kinds of topological defects, various substitutes
added into the structure, external influences such as a magnetic field, border effects given by the geometry
close to the defects, etc. In the case of the cone, the geometry is given by the pentagons in the tip whose
number changes from $1$ to $5$ and it determines the resulting vortex angle. The properties of nanocones are
widely investigated in \cite{lammertcrespi}. In this paper, we will be interested in how the border effects
influence the electronic structure close to the tip.

We apply the approximation used for the studies of the electronic
properties of the double wall nanotubes in \cite{pincak1}: we
consider the influence of the radius of the nanoparticles on the
bond hybridization and also on the $\pi$ orbitals. The
rehybridization of the $\sigma, \pi$ orbitals method was used in
order to compute the influence of the curvature of the nanocone shell
on the matrix elements. The electronic structure of the cone where
this effect is not considered was investigated in \cite{sitenko}.
The continuum gauge field-theory model was used, where the
disclination is represented by a SO(2) gauge vortex.
The solution of the Dirac-Weyl equation gives the local density of states ($LDoS$) around the Fermi energy.\\

In the continuum gauge field-theory, the following equation is solved to investigate the electronic structure of the conical surface:
\begin{equation}\label{first}\hat{H}\psi=E\psi.\end{equation}
Because of the atomic structure of the graphene lattice which is composed of two sublattices $A$ and $B$, the solution has
two components, $\psi=(\psi_A, \psi_B)$ which depend on the energy. From the solution, $LDoS$ is calculated as
\begin{equation}\label{LDoS}LDoS(E)=|\psi_A(E)|^2+|\psi_B(E)|^2.\end{equation}
In \cite{sitenko}, the Hamiltonian in (\ref{first}) has the form
\begin{equation}\label{second}\hat{H}=\left(\begin{array}{cc}H_1 & 0\\0 & H_{-1}\end{array}\right)\end{equation}
and
\begin{equation}\label{third}\hat{H}_s=\hbar v\left\{i\sigma^2\partial_r-\sigma^1r^{-1}\left[(1-\eta)^{-1}\left(is\partial_{\varphi}+\frac{3}{2}\eta\right)+
\frac{1}{2}\sigma^3\right]\right\},\hspace{5mm}s=\pm 1.\end{equation}
The Hamiltonian in (\ref{second}) arised from a Hamiltonian $H_0$ by performing some unitary transformations,
so we cannot strictly say to which of the definite sublattices $A$ or $B$ or the Fermi points "$+$" or "$-$"
the particular components of the Hamiltonian $H_+, H_-$ correspond which mix up different possibilities.
The parameters $r$ and $\varphi$ in (\ref{third}) represent the polar coordinates on the curved conical
surface with the origin in the tip, and $\hbar$ and $v$ represent the Planck constant and the Fermi velocity whose product $\hbar v=\frac{3}{2}t$ \cite{sitenko}, where $t\doteq 3\,{\rm eV}$ is the hopping integral corresponding to the neighbouring sites. The Pauli matrices $\sigma^i, i=1,2,3$ are a regular part of the Dirac-Weyl equation. The number of pentagonal defects $N_d$ is included in the parameter $\eta=N_d/6$.

In this paper, we will use the upgraded version of the Hamiltonian in (\ref{third}):
\begin{equation}\label{fourth}\hat{H}_s=\hbar v\left\{i\sigma^2\partial_r-\sigma^1r^{-1}\left[(1-\eta)^{-1}\left(is\partial_{\varphi}+\frac{3}{2}\eta\right)+
\frac{1}{2}\sigma^3\right]-\frac{A}{(1-\eta)^2r^2}I\right\},\end{equation}
so we supply a nonzero dimensionless parameter $A$ which represents the
dependence of the $\pi$ orbital on the local curvature of the
systems. And so we have (see Appendix)
\begin{equation}\label{2a}
\langle \pi |\hat{H}|\pi\rangle =\langle p_{z} |\hat{H}|p_{z}\rangle
- \frac{\hbar v A}{(1-\eta)^2r^2}
\end{equation}

Similar effect has been taken into account to compute the
electronic structure of the multiwalled fullerenes and nanotubes
in \cite{pincak1} and \cite{pincak2}. This curvature-dependent effect has nothing to do with the overlap of the $\pi$ orbitals.\\

\begin{figure}
\begin{center}
{\includegraphics[width=60mm]{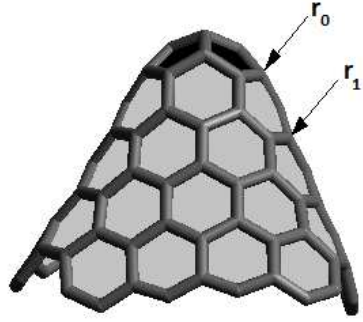}}\caption{The structure of the graphitic cone with 3 (pentagonal) defects.}\label{fg7}
\end{center}
\end{figure}

In Fig. \ref{fg7}, a model of the graphitic nanocone with 3 (pentagonal) defects in the tip is displayed. Two significant positions with respect to the tip ($r_0$ and $r_1$)
are marked there: $r_1$ denotes the upper limit of $r$ for which the mentioned effect of the pseudopotential is significant. On the other hand, at the distance $r_0$ from the tip,
the surface becomes too smooth and the influence of the $\pi$ bonds is negligible in comparison with the influence of the $\sigma$ bonds.
So only in the interval $(r_0, r_1)$ the investigation of the pseudopotential (or any kind of effective charge) has meaning. Denoting by $d$ the length of the $C-C$ bond, we can estimate $r_0\sim d, r_1\sim(2-3)d$ (here we choose $r_1=2d$).\\

If we write (\ref{first}) as a system of equations (with the Hamiltonian including the nonzero parameter $A$ presented in (\ref{fourth})), then, by performing the substitution method, after final corrections and putting
\begin{equation}\frac{A}{(1-\eta)^2}=a_1,\hspace{1cm}\frac{E}{\hbar v}=a_2,\hspace{1cm}\frac{sn-\eta}{1-\eta}=a_3,\end{equation}
we obtain
\[\left[a_2+\frac{a_1}{r^2}\right]\frac{{\rm d}^2f}{{\rm d}r^2}+\left[\frac{a_2}{r}+\frac{3a_1}{r^3}\right]\frac{{\rm d}f}{{\rm d}r}+\]
\begin{equation}\label{solve1}
+\left[a_2^3+(3a_1a_2-a_3^2)\frac{a_2}{r^2}+
(3a_1a_2-2a_3-a_3^2)\frac{a_1}{r^4}+\frac{a_1^3}{r^6}\right]f(r)=0.\end{equation}
Here, $f(r)$ represents an arbitrary component of the two-component wave function $\psi$ from (\ref{first}).\\

To solve the equation for low values of the parameter $r$, we use the Runge-Kutta numerical method. In this method, we suppose some initial values of the solution and its derivation at the point $r_0$:
\begin{equation}f(r_0)=f_0,\hspace{1cm}f'(r_0)=f_0'.\end{equation}
Then, we calculate the values of the given function at the points $x_0+n h$, where $h>0$ is a small number. At each step, we calculate the coefficients
\begin{equation}k_1(h)=\frac{1}{2}h^2f(x_0,y_0,y_0'),\end{equation}
\begin{equation}k_2(h)=\frac{1}{2}h^2f(x_0+\frac{h}{2},y_0+\frac{h}{2}y_0'+\frac{k_1}{4},y_0'+\frac{k_1}{h}),\end{equation}
\begin{equation}k_3(h)=\frac{1}{4}h^2f(x_0+\frac{h}{2},y_0+\frac{h}{2}y_0'+\frac{k_2}{4},y_0'+\frac{k_2}{h}),\end{equation}
\begin{equation}k_4(h)=\frac{1}{2}h^2f(x_0+h,y_0+h y_0'+k_3,y_0'+\frac{2k_3}{h}),\end{equation}
so that we finally find
\begin{equation}y(x_0+h)\doteq y(x_0)+h y'(x_0)+\frac{1}{3}(k_1(h)+k_2(h)+k_3(h)),\end{equation}
\begin{equation}y'(x_0+h)\doteq y'(x_0)+\frac{1}{3h}(k_1(h)+2k_2(h)+2k_3(h)+k_4(h)).\end{equation}
The detailed description of this method is given in \cite{runge-kutta}. We can try to find the analytical solution as well. In this case (low values of $r$),
this solution can be approximated by simplifying (\ref{solve1}) in the way that we suppose small values of $r$. Then, we get
\begin{equation}\frac{{\rm d}^2f}{{\rm d}r^2}+\left(\frac{3}{r}-\frac{2a_2}{a_1}r\right)\frac{{\rm d}f}{{\rm d}r}+f(r)=0\end{equation}
with the solution
\begin{equation}f_0(r)=C_1F_-(r)+C_2F_+(r)=C_1\hspace{1mm}    {}_1F_1\left(-\frac{a_1}{4a_2},2,\frac{a_2}{a_1}r^2\right)+C_2G_{1,2}^{2,0}\left(-\frac{a_2}{a_1}r^2\Big|
\begin{tabular}{cc}\multicolumn{2}{c}{$1+\frac{a_1}{4a_2}$}\\$-1$ & $0$\end{tabular} \right),\end{equation}
where  ${}_1F_1$ is the hypergeometric function, $G_{1,2}^{2,0}$ is the Meier $G-$function, $C_1, C_2$ are the normalization constants whose form will be introduced bellow.
In practical calculations, we will not consider the contribution of the Meier  $G-$function which will be very small. Then, $LDoS$ will have the form (taking in mind the meaning of the particular parameters)
\[LDoS(\tilde{E},\eta)=C_1(\tilde{E},\eta)^2{}_1F_1^2\left(-\frac{A}{4(1-\eta)^2\tilde{E}},2,
\frac{\tilde{E}}{A}(1-\eta)^2r^2\right)\sim\]
\begin{equation}\label{hyperF}
\sim C_1(\tilde{E},\eta)^2\left[\frac{1}{64}r^4+\left(\frac{\tilde{E}}{48}-\frac{A}{192(1-\eta)^2}\right)r^6\right],\end{equation}
where $\tilde{E}=a_2=\frac{E}{\hbar v}$. The reliability of this method of finding the solution is limited: without satisfying additional requirements on the Taylor expansion of the solution, the deviation  from the real solution can be considerable. We will verify the validity of this solution in the plots of $LDoS$.\\

For the high values of the $r$ parameter, we will simplify equation
(\ref{solve1}) by estimating which terms will be so negligible that
we can exclude them. After doing this, the form of the equation
(\ref{solve1}) for the high values of the $r$ parameter is
\begin{equation}\frac{{\rm d}^2f}{{\rm d}r^2}+\frac{1}{r}\frac{{\rm d}f}{{\rm d}r}+\left[a_2^2+\frac{1}{r^2}(3a_1a_2-a_3^2)\right]f(r)=0,\end{equation}
and the solution has the form
\begin{equation}\label{yinfty}f_{\infty}(r)=C_1G_-(r)+C_2G_+(r)=C_1 J_{-\frac{a_3}{a_2}}(\sqrt{r})+C_2 J_{\frac{a_3}{a_2}}(\sqrt{r}),\end{equation}
where $J$ is the Bessel function of the first kind. As there is no reason to prefer the site of a concrete type, we will suppose $C_1=C_2$. Then, the normalization constant will be calculated from the condition \begin{equation}C_1^2\int\limits_0^{r_{max}}(|G_-(r)|^2+|G_+(r)|^2){\rm d}r=1,\end{equation}
where $r_{max}$, the upper limit of the integration, will be chosen as $r_{max}=50$. This normalization constant will be used for the case of the low as well as the high values of $r$.\\



\begin{figure}
{\includegraphics[width=130mm]{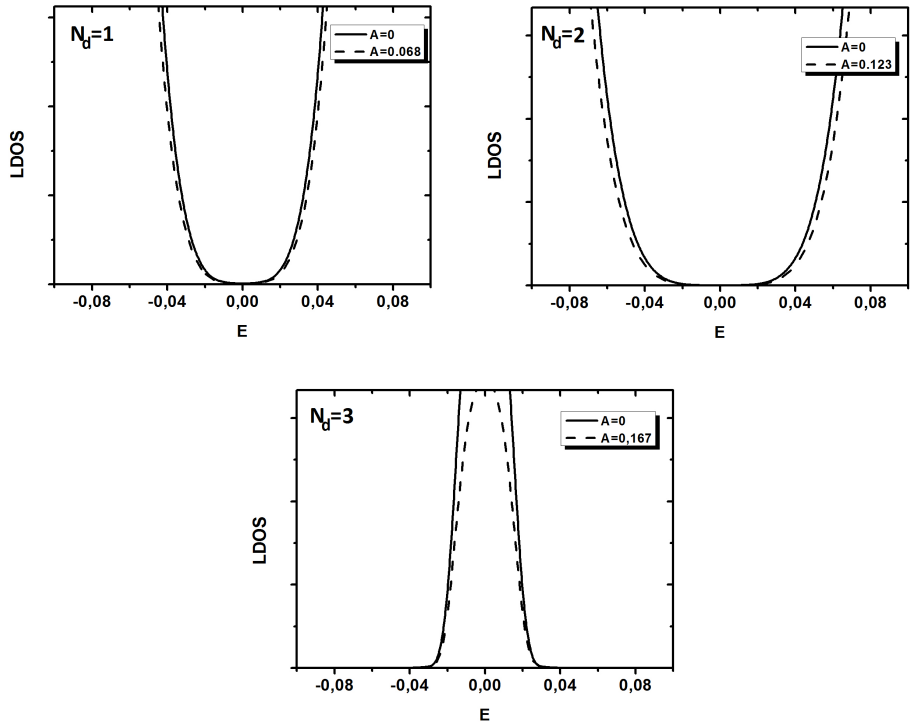}}\caption{Schematic plot of $LDoS$ for the wave function $f_0$ and low values of $r$ and comparison with the case without border effects for the number of pentagons $N_d=1, 2$ and $3,\hspace{2mm}n=2, s=1, r=2$.}\label{fg5}
\end{figure}

Now, the dependence of $LDoS$ on the energies is found with the help of (\ref{LDoS}) (and by using the normalization constant $C_{\infty}$). For the low values of $r$, it is shown in the plot in Fig. \ref{fg5}. The case of a different number of pentagons with $A$ being zero and nonzero is studied there. The exact value of $A$ is calculated in Appendix. We see that for nonzero $A$, $LDoS$ decreases and metallization appears for a higher number of pentagons. The plots are valid for the solution calculated by using the Runge-Kutta method as well as for the solution presented by the hypergeometric function (\ref{hyperF}).\\

For the long distance from the tip where we can neglect the term $A/((1-\eta)^{2}r^{2})$, $LDoS$ corresponds to the solution presented in \cite{sitenko} and to the approximation for high values of $r$ presented in \cite{lamcresp}. We can try to find the energy levels using the analogy with the calculations presented in \cite{berry}. There, the solution has the form of the Bessel function $J_l(k_{nl})$ and the equality is derived
\begin{equation}J_l(k_{nl})=J_{l+1}(k_{nl}).\end{equation}
We can use the solution and the approximation from
\cite{sitenko} and do the same procedure. Then,
\begin{equation}J_{\nu}\left(\frac{E}{\hbar v}r\right)=J_{\nu+1}\left(\frac{E}{\hbar v}r\right)\end{equation}
holds, where $\nu=\frac{sn-\eta}{1-\eta}$. From \cite{lamcresp} follows
\begin{equation}J_{\nu}\left(\frac{E}{\hbar v}r\right)\sim\sqrt{\frac{2\hbar v}{\pi Er}}\cos\left[\frac{Er}{\hbar v}-\left(\nu+\frac{1}{2}\right)\frac{\pi}{2}\right],\end{equation}
so, after the substitution, we get
\begin{equation}\frac{E}{\hbar v}=\frac{(m+0.5)\pi+sn}{r},\hspace{1cm}m\in\mathcal{Z}.\end{equation}

\section{Conclusion}

We studied the electronic properties of the graphitic nanocone near the tip. However, we found that it is very difficult to find the analytical solution. So we divided the problem into two cases: the case of low values of $r$ and the case of high values of $r$. For the high values of $r$, the solution is in fact the same as in the case considered in \cite{sitenko} which does not reflect the pseudopotential; for the low values of $r$, we used the numerical methods of finding the solution and we compared it with the analytical solution which was calculated from the simplified version of equation (\ref{solve1}) for the given case. In fact, this simplification was carried out without additional requirements on the form of the solution; nevertheless, the results appeared to be in agreement with the numerical solution. It follows from the plots in Fig. \ref{fg5} that for the case of 1 and 2 pentagons, the electron flux in the nanocone tip is spread over all interval of energies and, on the other hand, it is concentrated around the Fermi level in the case of 3 pentagons.

For the high values of $r$, the energy levels were calculated with the help of the method used in \cite{berry}. For the low values of $r$, the calculation of the energy levels is complicated because the found approximation of the analytical solution has not the form of the Bessel function. Furthermore, it follows from the plots of $LDoS$ in Fig. \ref{fg5} that there are no peaks, and it significantly restricts the possibility of the calculation of the energy levels from the numerical or analytical approximations of the solution.

We expected that the border effects close to the tip are caused by the curvature of the $\pi$ orbitals of the neighboring atoms. In a closer approximation, the hopping terms could be supplied into (\ref{fourth}) which are caused by the overlap of the neighbouring $\pi$ orbitals and also by the overlap of the $\pi$ orbitals corresponding to the sites located at the opposite sides from the tip. Both effects would be stronger in the case of adding next pentagons to the tip which creates the form of the carbon nanohorn. This could be considered in further calculations.

The localization of the electrons on the nanocone tip in the case of 3 pentagonal defects could be applied in the field of the electron microscopes. One of the papers dealing with this subject is \cite{chenchen}. There, the carbon nanocone is recommended as a good material for the probe tip in the atomic force microscopy because of its good physical and chemical properties in the comparison with other materials like silicon. A process of the tip fabrication via the E-beam induced and the chemical vapor deposition is suggested there. Another applicability in the field of the scanning tunneling microscopy is given in \cite{carbon}.\\

\section*{Acknowledgments}
The work was supported
by the Slovak Academy of Sciences in the framework of CEX NANOFLUID,
and by the Science and Technology Assistance Agency under Contract
No. APVV-0509-07 and APVV-0171-10, VEGA Grant No. 2/0037/13 and Ministry
of Education Agency for Structural Funds of EU in frame of project
26220120021, 26220120033 and 26110230061. R. Pincak would like to thank the
TH division in CERN for hospitality.

\appendix

\section{The computation of local $\pi$ orbital energy on the cone}

The geometry of the cone can be expressed in $(r,\varphi)$
parametrization as
\begin{equation}
x=\alpha r \cos\varphi, y=\alpha r \sin\varphi, z=\frac{c}{b}\alpha
 r
.\end{equation}
Here, $\alpha =\frac{b}{\sqrt{b^{2}+c^{2}}}$ and $b$,$c$ are
parameters which define the cone geometry. It can be shown that $\alpha
=1-\eta $, where $2\pi\eta$ is the deficit angle characterizing the
magnitude of the removing sector. In this cone parametrization the
squared length element on the cone takes the form
$ds^{2}=dr^{2}+(1-\eta)^{2}r^{2}d\varphi^{2}$. Now we have to
compute the local curvature of the cone, which determines the local
hybridization of the $p_{x}$, $p_{y}$, $p_{z}$ and $2s$ orbital on
the carbon atoms and so determines the local $\pi$ orbital. The local
curvature can be computed from the curvature tensor
$b^{\alpha}_{\beta}$(\cite{Frankel}). In our case, the only nonzero
matrix element of the curvature tensor is $b_{\varphi}^{\varphi}$. We
have

\begin{equation}
\label{a1} b_{\varphi}^{\varphi}=\frac{1}{R}=\frac{c}{b r},
\end{equation}
where $R$ is the local radius of the nanoparticle surface. Using the
formula derived in works (\cite{pincak1,pincak2})
\begin{equation}
\langle \pi |\hat{H}|\pi\rangle =\langle p_{z} |\hat{H}|p_{z}\rangle
- \frac{A_{0}}{R^2},
\end{equation}
where $R$ is local radius of the surface measured in $d$ unit ($d$ is
the bond length between carbon atoms) and $A_{0}\sim 1$ eV for
armchair and zig-zag nanotube. Now using the Eq.(\ref{a1}) we get
\begin{equation}
\langle \pi |\hat{H}|\pi\rangle =\langle p_{z} |\hat{H}|p_{z}\rangle
- \frac{c^{2} A_{0}}{b^{2} r^2},
\end{equation}
which can be expressed in the form
\begin{equation}\label{a3}
\langle \pi |\hat{H}|\pi\rangle =\langle p_{z} |\hat{H}|p_{z}\rangle
- \frac{\eta (2-\eta) A_{0}}{(1-\eta)^{2} r^2}.
\end{equation}
And so the constant $A$ in Eq.(\ref{2a}) has the form
\begin{equation}\label{a4}
A =\frac{\eta (2-\eta) A_{0}}{\hbar v}.
\end{equation}
We can see that in the case that there are no any defects (flat graphene sheet) and $\eta=N_d/6=0$, the pseudopotential $A=0$ is also zero.\\

\end{document}